\documentclass[12pt]{article}

\usepackage{amsmath}
\usepackage{amssymb}
\usepackage{epsfig}
\usepackage{graphicx}
\usepackage{cite}
\usepackage{multirow}
\usepackage{longtable}
\usepackage{lscape}
\usepackage{dcolumn}
\usepackage[dvips]{color}
\usepackage{rotating}

\allowdisplaybreaks[4] 

\makeatletter
\renewcommand{\fnum@table}{\textbf{\tablename~\thetable}}
\renewcommand{\fnum@figure}{\textbf{\figurename~\thefigure}}
\makeatother

\newcounter{myenumi}

\renewcommand{\themyenumi}{\roman{myenumi}}
{\end{list}}

\newlength{\myem}
\newcounter{mysubequation}[equation]

\makeatletter
\renewcommand{\section}{\@startsection{section}{1}{0em}{-\baselineskip}%
{\baselineskip}{\normalfont\large\bfseries}}
\renewcommand{\subsection}%
{\@startsection{subsection}{2}{0em}{-0.7\baselineskip}%
{0.7\baselineskip}{\normalfont\bfseries}}
\makeatother

\newcommand{\bi}{\begin{itemize}}
\newcommand{\ei}{\end{itemize}}

\newcommand{\be}{\begin{equation}}
\newcommand{\ee}{\end{equation}}
\newcommand{\bea}{\begin{eqnarray}}
\newcommand{\eea}{\end{eqnarray}}

\def\ket#1{| \,#1\, \rangle}

\def\scx#1#2{\langle \,#1\, |\, #2\, \rangle}

\newcommand\poi{Poincar$\acute{\rm e}$~}

\begin{document}

\begin{titlepage}

\renewcommand{\thefootnote}{\alph{footnote}}

\vspace*{-3.cm}
\begin{flushright}
\end{flushright}


\renewcommand{\thefootnote}{\fnsymbol{footnote}}
\setcounter{footnote}{-1}

{\begin{center} {\LARGE \sf Reply to the comment on \\ ``Topological
phase in two flavor neutrino oscillations"}
\end{center}}

\renewcommand{\thefootnote}{\alph{footnote}}

\vspace*{.8cm}
\vspace*{.3cm}
{\begin{center} {\large{\sc
                Poonam~Mehta\footnote[1]{\makebox[1.cm]{Email:}
                poonam@rri.res.in}
                }}
\end{center}}
\vspace*{0cm}
{\it
\begin{center}
Raman Research Institute, C. V. Raman Avenue,
\\  Bangalore 560 080, India.

\end{center}}

\vspace*{1.5cm}

\begin{center}
{\Large \today}
\end{center}

{\Large \bf
\begin{center} Abstract \end{center}  }
 In a recent paper, we showed that there is a neat
geometric interpretation of two flavor neutrino oscillation
formulae, and that the geometric phase involved in the physics of
oscillations is restricted to be topological as long as CP is
conserved. This paper has been criticised by Bhandari. In the
present note, we show that the criticisms are not valid and only
reflect his failure to understand some crucial points.

\vspace*{.5cm}

\end{titlepage}

\newpage

\renewcommand{\thefootnote}{\arabic{footnote}}
\setcounter{footnote}{0}


In Ref.~\cite{pmcpc}, we have  shown that the coherent phase
involved in  the neutrino oscillation formalism can be decomposed
into two parts, one corresponding to the usual dynamical phase and
the other part being geometric.  This study was carried out for the
specific case of CP conserving two flavor neutrinos. In this context
we noted that the geometric phase is actually
topological~\footnote{By our definition, the topological phase
refers to phase factors that are insensitive to small changes in the
circuit (and are invariant under deformations of circuit), while
geometric phases are sensitive to such changes.}.

The author of the comment~\cite{rb} claims to give a simple
derivation of the main result obtained in~\cite{pmcpc} and then goes
on to make some strong, unjustified and inappropriate remarks
towards the end. We disagree with all of what is claimed
in~\cite{rb} and in what follows, we rebutt these claims.

First let us address the scientific content of the
comment~\cite{rb}. The author of~\cite{rb} repeats the derivation of
the two flavor neutrino oscillation formulae as presented in our
paper~\cite{pmcpc}. There is no disagreement till Eq.~(4)
in~\cite{rb}. In Ref.~\cite{pmcpc}, we note that  the individual
interference terms (with the dynamical phase removed) appearing in
the expressions for survival ($c_+^\star c_- A_{\alpha}$ in the
notation of~\cite{rb}) and appearance ($c_+^\star c_- A_{\beta}$ in
the notation of~\cite{rb}) probabilities can be written as $r
e^{i\ss}$ where $\ss$ is the Pancharatnam phase. We find that this
phase can only take values $0$ or $\pi$ for CP conserving
interactions.

The disagreement appears because the author of~\cite{rb} is
computing   the relative phase between the interference terms of
survival (Eq.~(3)) and appearance (Eq.~(4)) probabilities i.e.
$A_{\beta}^\star A_{\alpha}$ (see Eq.~(6) in~\cite{rb}) which
{\it{differs}} from our quantity of interest. It is therefore no
surprise that the author of~\cite{rb} gets a different answer: his
phase is always $\pm \pi$. This result follows easily from
unitarity, but is {\it irrelevant to our discussion}. His final
expression Eq.~(6) (in~\cite{rb})
 only contains {\it{four}} states
 ($\ket{\nu_\alpha},\ket{\nu_\beta},\ket{\theta_2,+},\ket{\theta_2,-}$)
 and not six
 ($\ket{\nu_\alpha},\ket{\nu_\beta},\ket{\theta_1,+},\ket{\theta_1,-},\ket{\theta_2,+},\ket{\theta_2,-}$)
  for the general case of varying density matter (see Eq.~(16)
 of~\cite{pmcpc}).

On page 3, the author of~\cite{rb} also  gives a connection of the
$\pm \pi$ phase in neutrino system with phase shifts observed in
polarisation optics. This correspondence is discussed in detail
in~\cite{pmcpc}. This analogy is precisely the reason for viewing
neutrino oscillations using ideas current in polarisation optics. We
would like to add that such analogous optics experiments to
demonstrate the physics involved in two flavor neutrino oscillations
have been carried out recently by Weinheimer~\cite{wein}. However
spatial split beam experiments which are common in optics are
impossible for neutrinos because of their low refractive index. To
get around this, we consider neutrino oscillations in terms of a
split-beam two-path interference experiment in {\it{energy space}}.
The author of~\cite{rb} missed this most crucial point
of~\cite{pmcpc}. From this viewpoint, the cross term in the survival
and appearance probability can be viewed as a series of quantum
collapses with intermediate adiabatic evolutions (in varying density
matter) which may or may not enclose a solid angle in the ray space.
The phase of the cross terms (with the dynamical phase removed) is
the Pancharatnam phase. Hence in~\cite{pmcpc}, we are only
interested in the individual cross terms appearing in the
probability and not in the relative phase as computed in~\cite{rb}.
 This settles the major
scientific objection raised in~\cite{rb}.

Next, we address the criticism  on page 3 of~\cite{rb} regarding
misleading statements made in our paper~\cite{pmcpc}. Before we
predict the value for the geometric phase~\footnote{The ``geometric
phase" is  by definition the phase that is not dynamical and it is
obtained after dropping the dynamical phase from the interference
term.} using Pancharatnam's ideas~\cite{Pancharatnam:1956}, we have
very clearly used the phrase {\it{``Upon removing the dynamical
phase"}} on page 7 (after Eq.~(16) in~\cite{pmcpc}) which the author
of~\cite{rb} missed and therefore was led to an incorrect
understanding of the paper. Hence our claim is correct. We would
like to stress that {\it{the dynamical phase is by no means ignored
in the paper~\cite{pmcpc}}} - it is present in all expressions
including the final equations (Eq.~(21) of~\cite{pmcpc}) when we
compare our result in the special case with standard expressions for
probability used in literature. The author of~\cite{rb} claimed
(wrongly of course) that geometric phase factors in the cross terms
in Eqs.~(3) and (4) are constant in time only for the case of
vacuum. Whereas, what we claimed was that the geometric part in the
total phase is always topological i.e. restricted to take values $0$
or $\pi$ irrespective of the properties of medium as long as we have
a CP conserving form of the total Hamiltonian which means it holds
for vacuum or any medium with either a constant or slowly varying
density. This is because for the CP conserving case, the states are
real and the cyclic loop formed by collapses of neutrino states is
always restricted to a great circle on the \poi sphere. The
topological phase depends on the actual winding around this
equatorial great circle and is either $0$ or $\pi$. Our result of
``$\pi$ anholonomy" is also consistent with the first available
papers (around 1958) on geometric phases in CP conserving two by two
case in the field of molecular physics~\cite{lh}.

Now we discuss the objection raised on page 3 in~\cite{rb} regarding
a sentence in the abstract :

{\it{``Our study shows for the first time that there is a geometric
interpretation of the neutrino oscillation formulae for the
detection probability of neutrino species."}}

We would like to mention that the full sentence in the abstract
of~\cite{pmcpc} is  actually the following (and not what is said in
the comment~\cite{rb})

{\it{``For the minimal case of two flavors and CP conservation, our
study shows for the first time that there is a geometric
interpretation of the neutrino oscillation formulae for the
detection probability of neutrino species."}}

We think this statement is apt because of the reasons that have
already been listed in detail in the paper~\cite{pmcpc}. We repeat
them here. Note that the statement is made   for the minimal case of
two flavors and CP conservation. The first point to note  is that in
two flavor case, we can have up to three independent parameters
appearing in the Hamiltonian. In case of vacuum or constant density
matter, there is no varying parameter while for varying density
matter, there is only one such parameter (the electron number
density). Hence appearance of cyclic Berry phase is ruled out. Of
course one can think of the generalized geometric phase and some
people claimed appearance of geometric phase in the two flavor case,
but it should be noted that such terms appeared at the level of
amplitudes and not probabilities. In~\cite{pmcpc}, we clarified that
due to the inherent limitation of designing a split-beam experiment
with neutrinos in the physical space, it was impossible to access
phases appearing at the level of amplitudes. So the existing
literature led to the belief that the formulae for two flavor
neutrino oscillation probability were devoid of any geometric phase
contribution. However, contrary to all the existing claims, our
study showed that there exists a topological contribution to the
total phase at the probability level even for the minimal case of
two flavors and CP conservation. And, to the best of our knowledge,
this was noted for the first time in~\cite{pmcpc}.

 Regarding the correctness of the following statement as questioned on page 3 in~\cite{rb} :

 {\it{``More precisely  the standard result for neutrino oscillations is in fact
  a realization of the Pancharatnam topological phase."}}

Because of the topological phase being inherently built into the
structure of the neutrino mixing matrix, it is true that the
standard formalism of oscillations is actually a realization of
Pancharatnam's phase. The minus sign appearing in the two by two
orthogonal rotation matrix has the interpretation of being the
Pancharatnam phase of $\pi$.

There were some general comments in~\cite{rb} concerning relevance
and usefulness of our paper to which we would like to respond. There
was some confusion among the neutrino community as to whether the
geometric phase is an extra contribution to be added to the standard
treatment. Our paper shows that the geometric phase is topological
(for CP conserving, two flavor case) and contained in the standard
treatment of neutrino oscillations. It therefore serves a useful
scientific purpose.

To counter the criticism in~\cite{rb} regarding use of ``high
sounding language" in~\cite{pmcpc} (Physical Review), we would like
to remark that the language used is quite accessible to the
readership of Physical Review and considerably simpler than for
example, that of the Letter~\cite{sam}, which is far more technical
and involves abstract differential geometry. This
criticism~\cite{rb} sounds very strange coming from a co-author on
the PRL paper!

Having thus disposed of the rather meagre scientific content of the
objections raised in~\cite{rb}, we now come to what appears to be
the main point of this manuscript : the question of priority. The
main claim made in~\cite{rb} is that the idea of using the
{\it{shortest geodesic rule}} was first put forward in~\cite{rb7}
and not in~\cite{sam}. However this can not be sustained and the
author has been caught in the unfortunate position of  having been
``scooped'' by himself.
 The author of~\cite{rb} objects to the ``use of term shorter along with
  the reference to~\cite{sam}" in our paper~\cite{pmcpc} by
claiming that the words ``shortest" or ``shorter" appear nowhere in
the formulation of the geodesic rule in the paper by Samuel and
Bhandari~\cite{sam}. This claim is patently false. The word
``shortest'' appears in footnote 11 on page 2342 of~\cite{sam} and
logically complete supporting argument given there. The footnote is
not only a part of the paper but very much a part of the proof of
the geodesic rule in question. We first reproduce the footnote
from~\cite{sam} which contains the word shortest (underlined below),

{\it{``In fact, $\scx{\tilde \phi (0)}{\tilde \phi (s)}$ is also
positive if $\ket{\tilde \phi (s)}$ is the {\underline{shortest}}
geodesic connecting $\ket{\tilde \phi (0)}$ with $\ket{\tilde \phi
(1)}$. Along this  curve, $\scx{\tilde \phi (0)}{\tilde \phi (s)}$
never vanishes and, since it was positive to start with, remains
so."}}

While the footnote is enough to convincingly dismiss the claim, we
would like to add the following. The author of the comment~\cite{rb}
is evidently confused  about the use of ``any" in the phrase ``any
geodesic arc" on page 2341 in his own paper~\cite{sam}. A reading of
the relevant paragraph shows that the discussion is carried out in $
\mathcal{N}$ the set of normalizable states, which admits gauge
freedom. The geodesic equation is written in  $ \mathcal{N}$ and is
gauge invariant. This gauge freedom means that gauge copies of
geodesics are also geodesics. ``Any" refers to any of the gauge
copies. The discussion of the footnote referred to earlier is in the
{\it{ray space}} and clearly states that it is the {\it{shortest
geodesic}} that is relevant.

 In closing we remark that the tone and language of~\cite{rb}
 are inappropriate for any scientific journal.
 The scientific content of the comment is meagre and the main point
of priority unsustainable~\cite{footnote}. We believe that such
unseemly and fruitless controversies over minute points of
priority~\cite{footnote} are a drain on the time of authors,
editors, referees and readers and do not belong in any scientific
journal.

\subsubsection*{Acknowledgments}
It is a pleasure to thank Joseph Samuel and Supurna Sinha for many
exciting  discussions and a careful reading of this manuscript.


\begin{thebibliography}{10}


\bibitem{pmcpc}
P.~Mehta, {\it Topological phase in two flavor neutrino
oscillations},  {\em
  Phys. Rev. D} {\bf 79}, 096013  (2009).

\bibitem{rb}
R.~Bhandari, {\it Comment on ``Topological phase in two flavor
neutrino oscillations"}, {\em arXiv:1007.5935v1 [hep-ph]}.


\bibitem{Pancharatnam:1956}
S.~Pancharatnam, {\it Generalized theory of interference and its
applications},
   {\em Proc. Ind. Acad. Sci.} {\bf A44}  247 (1956).

\bibitem{wein} C.~Weinheimer, {\it Neutrino oscillations with a polarized laser beam: an analogical
demonstration experiment}, {\em arXiv.org:1001.2749 [quant-ph]}.


\bibitem{lh}
H.~C. Longuet-Higgins, U.~Opik, M.~H.~L. Pryce, and R.~A. Sack, {\it
Studies of
  the jahn-teller effect ii. the dynamical problem},  {\em Proc. Roy. Soc.
  Lond.} {\bf A244}, 1 (1958).


\bibitem{rb7}
R. Bhandari,  {\it{Observable Dirac-Type Singularities in Berry's
Phase and the Monopole}}, {\em Phys.
  Rev. Lett.} {\bf 88}, 100403 (2002).

\bibitem{sam}
J.~Samuel and R.~Bhandari, {\it General setting for berry's phase},
{\em Phys.
  Rev. Lett.} {\bf 60}, 2339  (1988).

\bibitem{footnote} Actually, the distinction between the greater and
shorter arc on the \poi sphere appears even earlier in papers by
Pancharatnam~\cite{Pancharatnam:1956}, Berry~\cite{berrypanc} and
possibly others!

\bibitem{berrypanc} M.~V.~Berry, {\it{The adiabatic phase and
Pancharatnam's phase for polarised light}}, {\em Jour. Mod. Opt.}
{\bf 34}, 1401 (1987).





\end{thebibliography}

\end{document}